\documentclass{cimento}

\usepackage{graphicx}
\usepackage{amsmath}
\usepackage{amssymb}
\usepackage{amsfonts}
\usepackage{bbm}
\usepackage[T1]{fontenc}
\usepackage[utf8]{inputenc}
\usepackage[english]{babel}
\usepackage{mathbbol}
\usepackage{xcolor}
\usepackage[normalem]{ulem}
\usepackage{cite}

\newcommand{\cR}{\mathcal{R}}

\newcommand{\p}{\partial}

\newcommand{\asg}{asymptotically safe gravity}
\newcommand{\Asg}{Asymptotically safe gravity}

\newcommand{\UV}{{\small{UV}}}
\newcommand{\IR}{{\small{IR}}}
\newcommand{\RG}{{\small{RG}}}
\newcommand{\FRGE}{{\small{FRGE}}}

\newcommand{\GN}{\ensuremath{G_N}}

\title{Form Factors in Quantum Gravity -- \newline contrasting  non-local, ghost-free gravity and Asymptotic Safety}

\shorttitle{Form Factors in Quantum Gravity}

\author{B.~Knorr\from{ins:x}\ETC,
C.~Ripken\from{ins:y}
\atque
F.~Saueressig\from{ins:z}
}
\instlist{\inst{ins:x} Perimeter Institute for Theoretical Physics \\ 31 Caroline Street North, Waterloo, ON N2L 2Y5, Canada
  \inst{ins:y} Institute of Physics (THEP) \\ University of Mainz, Staudingerweg 7, 55128 Mainz, Germany
\inst{ins:z} Institute for Mathematics, Astrophysics and Particle Physics (IMAPP) \\ Radboud University Nijmegen, Heyendaalseweg 135, 6525 AJ Nijmegen, The Netherlands}

\begin{document}

\maketitle

\begin{abstract}
Form factors constitute the key building block when organising the gravitational dynamics in terms of a curvature expansion. They generalise the concept of momentum-dependent couplings to curved spacetime. Moreover, they may capture modifications of the gravitational interactions at high energy scales in a non-perturbative way. We briefly review the role of form factors in various quantum gravity programmes  with special emphasis on non-local, ghost-free gravity and \asg{}. In this context, we introduce a quantum gravity motivated scalar toy model, which demonstrates that Lorentzian asymptotic safety may be compatible with the causal propagation of gravitons.
\end{abstract}

\section{Introduction}
Relativistic quantum field theories provide the gold standard for describing the  interactions of elementary particles mediated by the electroweak and strong force. In particular, the Standard Model of Particle Physics predicts the outcome of collider experiments and precision measurements with uncanny accuracy. This raises the intricate question whether gravity may also be consistently formulated as a relativistic quantum field theory. This inquiry may serve as a catalyst for sharpening our understanding of the fundamental principles followed by nature. In particular, notions like unitarity, positivity, and causality, which are often taken as fundamental when  dealing with particle physics in a fixed Minkowski spacetime, may acquire a different status when spacetime becomes a dynamical (and even fluctuating) entity.

Applying the perturbative quantisation techniques developed in the context of the Standard Model of Particle Physics to General Relativity leads to problems though. Owed to the negative mass dimension of Newton's coupling, this results in a perturbatively non-renormalisable quantum field theory. This result has sourced a broad range of research programmes striving for the reconciliation of gravity with quantum mechanics. Within the framework of relativistic quantum field theory, prominent candidates include quadratic gravity~\cite{Stelle:1976gc,Salvio:2018crh}, Conformal Gravity~\cite{Mannheim:2011ds}, non-local (ghost-free) gravity~\cite{Modesto:2017sdr,Buoninfante:2020ctr}, \asg{}~\cite{Percacci:2017fkn,Reuter:2019byg,Eichhorn:2018yfc,Pawlowski:2020qer}, and Causal Dynamical Triangulations~\cite{Ambjorn:2012jv,Loll:2019rdj}. A common element of these constructions is that they modify the gravitational dynamics at short distances.

Conformal Gravity utilises a bare action given by the square of the Weyl tensor. Unitarity may then be saved by noticing that the theory is a {\small{PT}}-symmetric theory, rather than a Hermitian theory. In a similar spirit, Quadratic Gravity introduces four-derivative terms in the gravitational action which make the theory perturbatively renormalisable (asymptotically free). At the same time, the new terms introduce additional, massive degrees of freedom which were long believed to be in conflict with unitarity. The recent interpretation of these degrees of freedom as ``Merlin particles'' propagating backward in time saves unitarity at the cost of violating micro-causality at scales associated with the mass of these particles \cite{Anselmi:2018tmf,Donoghue:2019fcb}.

More elaborate constructions in the realm of non-local gravity theories introduce a non-locality scale $\ell_{\rm non-local}$ and an additional momentum dependence in the propagators and vertices, which becomes operative on scales $\ell \lesssim \ell_{\rm non-local}$. These lead to an exponential~\cite{Modesto:2011kw,Biswas:2011ar} or polynomial~\cite{Modesto:2017sdr} suppression of the graviton propagator at high energies.\footnote{For earlier applications of non-local gravity models in cosmology see \cite{Biswas:2005qr,Deser:2007jk}.}  If the non-local modifications come in the form of entire, analytic functions, the construction also does not introduce new degrees of freedom, in variance with Quadratic Gravity. Specific models then arise from specifying the entire functions which (when exponentiated) appear in the propagators of the theory. In this way one arrives at a large class of (super-)renormalisable gravity models which deviate in their predictions for the physics at and above the non-locality scale.

\Asg{} builds on a formulation which is similar to the one used in non-local gravity theories. The key idea is that physical quantities like scattering amplitudes stay finite at very high energies. For gravity, this requires a delicate interplay between the propagators and the vertices capturing the interactions. Within \asg{}, the necessary relations are expected to arise from a renormalisation group fixed point associated with an interacting theory. At this fixed point, the theory enjoys an enhanced symmetry, so-called quantum scale invariance \cite{Wetterich:2019qzx}. The low energy behaviour is obtained by breaking this symmetry, and having a renormalisation group flow into a classical regime where effective field theory provides a good approximation.

The goal of these proceedings is to step outside of the traditional \asg{} discussion formulated in terms of functional renormalisation group equations and  flows, and assess the programme and its findings from the unifying perspective provided by the form factor formulation of the gravitational effective action. This will allow us to make contact with  other approaches to quantum gravity rooted within the realm of relativistic quantum field theory.

\section{\Asg{} in a nutshell}
\label{sect.2}
\Asg{} strives to construct a consistent and predictive quantum theory for gravity and gravity-matter systems. In this section, we will review the key elements in this programme, highlighting possible areas of overlap with non-local, ghost-free gravity.

\subsection{The conceptual framework}
\label{sect2.1}
Newton's coupling \GN{}, determining the strength of the gravitational interactions in a four-dimensional spacetime, comes with a negative mass dimension $[\GN{}] = -2$. A direct consequence of this peculiar property is that scattering amplitudes related to gravity-mediated scattering events in General Relativity develop divergences at high energies when computed perturbatively using standard Feynman diagram techniques. The order of the divergences grows with the order of the perturbation theory, signaling that the  result is trustworthy below a certain \UV{}-scale $\Lambda_{\rm UV}$ only.
The central idea of \asg{} is to cure these divergences, rendering the (physical) correlation functions \emph{finite} and \emph{scale-free} at trans-Planckian energy scales. This point makes \asg{} qualitatively different from the philosophy underlying the non-local, ghost-free gravity constructions, where these amplitudes are supposed to vanish exponentially at energies exceeding the non-locality scale.

 Technically, the high-energy behaviour of an asymptotically safe theory is controlled by  an interacting fixed point of the theory's renormalisation group flow (the Reuter fixed point).  At such a fixed point, quantum fluctuations balance out the classical mass dimensions of the coupling constants in such a way that the dimensionless couplings entering  physical observables remain finite.
The low-energy physics visible from the effective field theory treatment of General Relativity emerges from ``flowing away from the fixed point'' into a region where physics is controlled by the free fixed point of the theory~\cite{Reuter:2001ag,Reuter:2004nx}. This crossover ensures that below a transition scale, quantum fluctuations are small, and one recovers classical physics to leading order. The predictive power of \asg{} arises from the condition that the flow has to emanate from the fixed point along a \UV{}-relevant direction. Typically, this leaves a low number of free parameters determining ``the direction'' in which the \RG{} flow exits the fixed point. The physics is then completely fixed by this set of parameters. Heuristically, the predictive power of asymptotically safe gravity may be understood as follows. For a local, perturbatively renormalisable quantum field theory, the number of free parameters is given by the power-counting relevant and marginal operators. For the fixed points investigated within asymptotically safe gravity the classical power counting receives quantum corrections. These corrections are not strong enough though to turn an arbitrary number of classically irrelevant couplings to relevant ones. This feature has led to the characterisation of the Reuter fixed point as an ``almost Gaussian'' fixed point \cite{Falls:2013bv}.\footnote{Results obtained within the composite operator formalism~\cite{Kurov:2020csd} suggest that gravity-matter systems may also possess genuinely non-Gaussian fixed points where the scaling of the couplings is dominated by quantum corrections in such a way that classical power counting does no longer serve as a good ordering principle. Since the identification of such fixed points is technically difficult, not much is known about this possibility at the present stage.}

A key element in the Asymptotic Safety programme is that the existence of a suitable renormalisation group fixed point is not an input -- it must be established based on explicit computations. The primary tool for determining the fixed points of gravity (potentially supplemented by additional matter degrees of freedom) is the functional renormalisation group equation (\FRGE{}) for the effective average action $\Gamma_k$~\cite{Wetterich:1992yh,Morris:1993qb,Reuter:1993kw} adapted to gravity~\cite{Reuter:1996cp} (also see \cite{Dupuis:2020fhh} for an up-to-date review and further references). By construction the \FRGE{} lives on Euclidean signature spacetimes and takes the explicit form
\begin{equation}
\label{eq:FRGE}
k \partial_k \Gamma_k = \frac{1}{2} {\rm Tr}\left[\left(\Gamma_k^{(2)} + \cR_k \right)^{-1} \, k \partial_k \cR_k \right] \, . 
\end{equation}
Here $k$ is the coarse-graining scale at which quantum fluctuations are integrated out, $\Gamma_k^{(2)}$ denotes the second variation of $\Gamma_k$ with respect to the fluctuation fields, and Tr contains a sum over fields and discrete indices as well as an integration over loop momenta. The regulator $\cR_k$ provides a mass-term for fluctuations with momenta $q^2 \lesssim k^2$ and ensures that the trace is free from \UV{} and \IR{} divergences.

The \FRGE{} allows to integrate out quantum fluctuations shell-by-shell in momentum space by lowering $k$. Conventionally, $k=0$ corresponds to the point where all fluctuations have been integrated out. As a consequence, $\Gamma_{k=0}$ agrees with the quantum effective action $\Gamma$. By definition, the latter contains the fully dressed propagators and vertices. Hence any physical process can be computed by a tree-level computation based on the vertices and propagators derived from $\Gamma$. Conceptual questions, including causality, unitarity, and positivity, should thus be analysed based on $\Gamma$. Note that even for the case where the bare action is strictly local, $\Gamma$ is known to contain non-local and also non-analytic terms. The prototypical example arises from the perturbative quantisation of the Einstein-Hilbert action in the framework of effective field theory~\cite{Donoghue:2017pgk}, where $\Gamma$ contains terms logarithmic in the momentum \cite{Percacci:2017fkn}.

Fixed points $\Gamma_*$ correspond to stationary points of \eqref{eq:FRGE}. Note that $\Gamma_*$ is not identical to the bare action $S^{\rm bare}$ appearing in the path integral. The bare action can be reconstructed from $\Gamma_*$ by solving the reconstruction problem~\cite{Manrique:2008zw,Morris:2015oca}, stating that the actions are related by a functional trace similar to the one appearing in \eqref{eq:FRGE}. As a direct consequence arguments related to causality and unitarity which are formulated at the level of the bare action do not necessarily carry over to the fixed point action $\Gamma_*$ and vice versa. In particular, a bare action which is manifestly local may be mapped to a non-local $\Gamma_*$.

Solutions of the \FRGE{} that interpolate between $\Gamma_*$ for $k \rightarrow \infty$ and $\Gamma$ at $k=0$ determine the quantum effective action in terms of the free parameters encoding ``the direction'' in which the solution emanates from $\Gamma_*$. It is expected that this provides significant predictive power for the physics compatible with Asymptotic Safety.

\subsection{Asymptotic Safety at the level of scattering amplitudes}
\label{sect2.2}
Two key questions related to the Asymptotic Safety construction described in the last subsection are
\begin{itemize}
	\item[a)] whether the (Lorentzian) quantum effective action $\Gamma$ associated with a given gravity-matter system \emph{has sufficient room} in order to  render all scattering amplitudes that are scale-free and finite in the \UV{}, and
	\item[b)]  are the structures that  realise this feature  compatible with requiring unitarity and causality?
\end{itemize}

Within the form factor approach to quantum gravity introduced in~\cite{Knorr:2019atm}, Ref.~\cite{Draper:2020bop} answered both of these points affirmatively. The key idea of the form factor programme is to expand the quantum effective action $\Gamma$ in powers of the spacetime curvature while retaining the full functional dependence on the spacetime derivatives acting on the curvatures. Loosely speaking, this generalises the coupling constants to functions of the  momenta while retaining covariance under general coordinate transformations. For metric gravity, when including the terms up to order $\cR^3$, this expansion reads
\begin{equation}
\label{eq:curvatureexp}
\Gamma[g] = \frac{1}{16\pi \GN{}} \int \text{d}^4x \sqrt{-g} \, \left[ 2\Lambda - R - \frac{1}{6} R f_R(\Delta) R + \frac{1}{2} C_{\mu\nu\rho\sigma} f_C(\Delta) C^{\mu\nu\rho\sigma} + \cdots\right] \, .
\end{equation}
Here $\Lambda$ denotes the cosmological constant and $\Delta = - g^{\mu\nu} D_\mu D_\nu$ is the d'Alembertian. The two form factors $f_R(\Delta)$ and $f_C(\Delta)$ generalise the notion of running couplings to the quantum effective action in a curved spacetime. In particular, they completely fix the graviton propagator in a flat spacetime. \hyphenation{an-a-lyse}
The perspective taken by the form factor programme is manifestly different from organising $\Gamma$ in terms of a derivative expansion, which retains interaction terms up to a finite mass-dimension only. While the latter constitutes a powerful approach to analyse low-energy physics in the spirit of an effective field theory, it is clear that the derivative expansion fails by construction, e.g., when trying to address questions related to the presence of ghosts in the gravitational propagators~\cite{Knorr:2019atm, Platania:2020knd}.

In order to describe gravity-mediated matter interactions, \eqref{eq:curvatureexp} must be supplemented by a suitable matter action. For the $\phi\phi \to \chi\chi$ processes described in~\cite{Draper:2020bop},
\begin{equation}
	\begin{split}
\label{eq:matter}
& \Gamma^{\text{matter}}[\phi,\chi,g] = \int \text{d}^4x \sqrt{-g} \, \Big[ \frac{1}{2} \phi f_\phi(\Delta) \phi + \frac{1}{2} \chi f_\chi(\Delta) \chi + f_{\phi\chi} \phi^2 \chi^2 \\ & \quad
 + f_{R\phi\phi} R \phi^2 + f_{Ric\phi\phi} R^{\mu\nu} (D_\nu D_\mu \phi) \phi +  f_{R\chi\chi} R \chi^2 + f_{Ric\chi\chi} R^{\mu\nu} (D_\nu D_\mu \chi) \chi + \cdots \Big] \, . 
\end{split}
\end{equation}
Here we introduced momentum-dependent form factors $f_\phi, f_\chi$ in the scalar kinetic terms as well as a form factor $f_{\phi\chi}$ in the scalar self-interaction. The non-minimal gravity-matter form factors are listed in the second line. All form factors depend on the momenta (in terms of covariant derivatives) of all fields they multiply. Eqs.\ \eqref{eq:curvatureexp} and \eqref{eq:matter} contain all terms that could potentially contribute to the  $\phi\phi \mapsto \chi\chi$-scattering process. The most general amplitude compatible with a relativistic quantum field theory can then be obtained from computing the corresponding tree-level Feynman diagrams based on the vertices derived from the quantum effective action. Note that specifying the amplitude completely requires the explicit form of the form factors \emph{in the gravitational and matter sector}.

A set of form factors exhibiting the characteristic properties of scattering amplitudes expected from \asg{} have been suggested in~\cite{Draper:2020bop} where
\begin{equation}\label{tanhff}
f_{C} = \Lambda_C^{-2} \tanh(\Delta/\Lambda_C^2) \, , \qquad f_{R} =\Lambda_R^{-2}  \tanh( \Delta/\Lambda_R^2) \, . 
\end{equation}
The parameters $\Lambda_{R,C}^2$ set the scale at which the resulting amplitude starts deviating significantly from the one obtained in General Relativity. From the perspective of \asg{}, it is natural to associate these scales with the scale separating the low-energy physics controlled by the free fixed point and the quantum regime described by the Reuter fixed point. Typically, one would then associate this scale with the squared Planck mass, even though experimental bounds would allow for much smaller values. 

 For the specific choice \eqref{tanhff} the resulting partial wave amplitudes related to the exchange of gravitons have been given in \cite{Draper:2020bop}. Their behaviour is qualitatively similar to the one shown in the left diagram of Fig.\ \ref{Fig.amp}.
The key feature of this analysis is that the amplitudes are rendered finite as the centre-of-mass energy goes to infinity. This effect is caused by an infinite tower of poles in the propagator, situated at purely imaginary momenta, as well as the interplay among the propagator and momentum-dependent self-interactions in the matter sector. The next section illustrates the key features of the construction in a scalar toy model.
 
\section{A scalar toy model for asymptotically safe amplitudes}
\label{sect.3}
We now introduce a scalar toy model in Minkowski spacetime that captures the essential features of asymptotically safe amplitudes within the form factor programme. For concreteness, consider a real, massless scalar field $\phi$ in Minkowski spacetime with the quantum effective action given by
\begin{equation}\label{non-locffaction}
\Gamma[\phi] = \frac{1}{2} \int_x \phi f(-\p^2) \phi + \frac{\lambda}{2} \int_x (\p^2 \phi) \,  \phi^2 \, . 
\end{equation}
 Here the momentum dependence in the scalar self-interaction has been chosen such that the coupling $\lambda$ has negative mass-dimension $[\lambda^2] = -2$, similar to Newton's coupling in gravity. In order to ease the notation, we will set the scale $\Lambda = 1$. 
 
We now use the action \eqref{non-locffaction} in order to contrast asymptotically safe gravity and non-local, ghost-free gravity at the level of gauge-invariant amplitudes. For this purpose, we introduce the asymptotic safety-inspired scalar-kinetic term
\begin{equation} \label{eq:fftanh2}
f^{\rm tanh}(x) = x \left[1+x  \tanh(x)\right] \, . 
\end{equation}
In addition, we also consider a specific form factor mimicking the exponential decay of a non-local, ghost-free (nl) gravity model \cite{Buoninfante:2020ctr}
\begin{equation}\label{eq:ffnon-local}
f^{\rm nl}(x) = x \, \left[  e^x \right] \, . 
\end{equation}
The momentum-space representations of the propagators resulting from the choices \eqref{eq:fftanh2} and \eqref{eq:ffnon-local} are then given by
\begin{equation}
G^{\rm tanh}(p^2) = \frac{i}{(p^2+i\epsilon)(1+p^2 \tanh(p^2))} \, , \qquad 
G^{\rm nl}(p^2) = \frac{i}{(p^2+i\epsilon)} e^{-p^2} \, . 
\end{equation}
Here we did not include an ``$i\epsilon$''-prescription in the tanh and the exponential, since there is no ambiguity in how the Feynman contour integral should include the poles in the complex plane.

\begin{figure}[t!]
	\includegraphics[width = 0.48\textwidth]{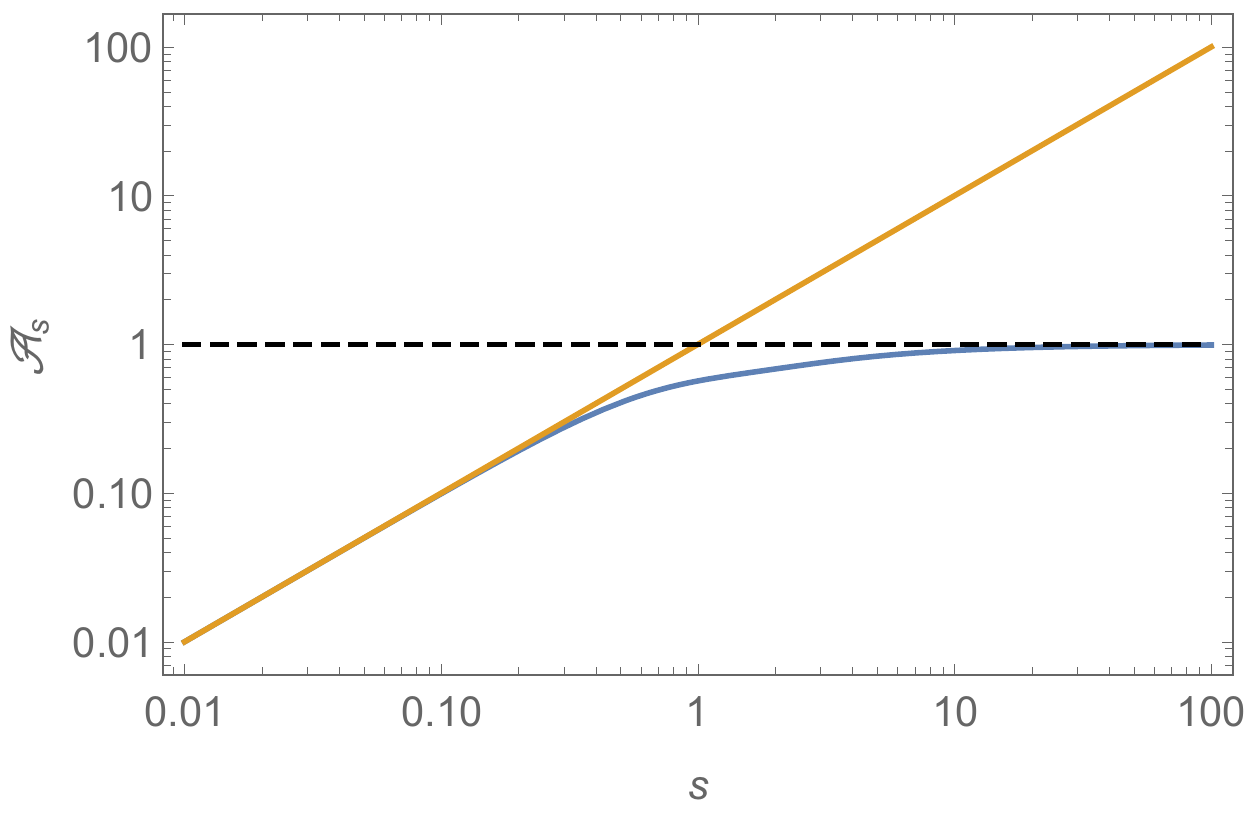} \;
	\includegraphics[width = 0.48\textwidth]{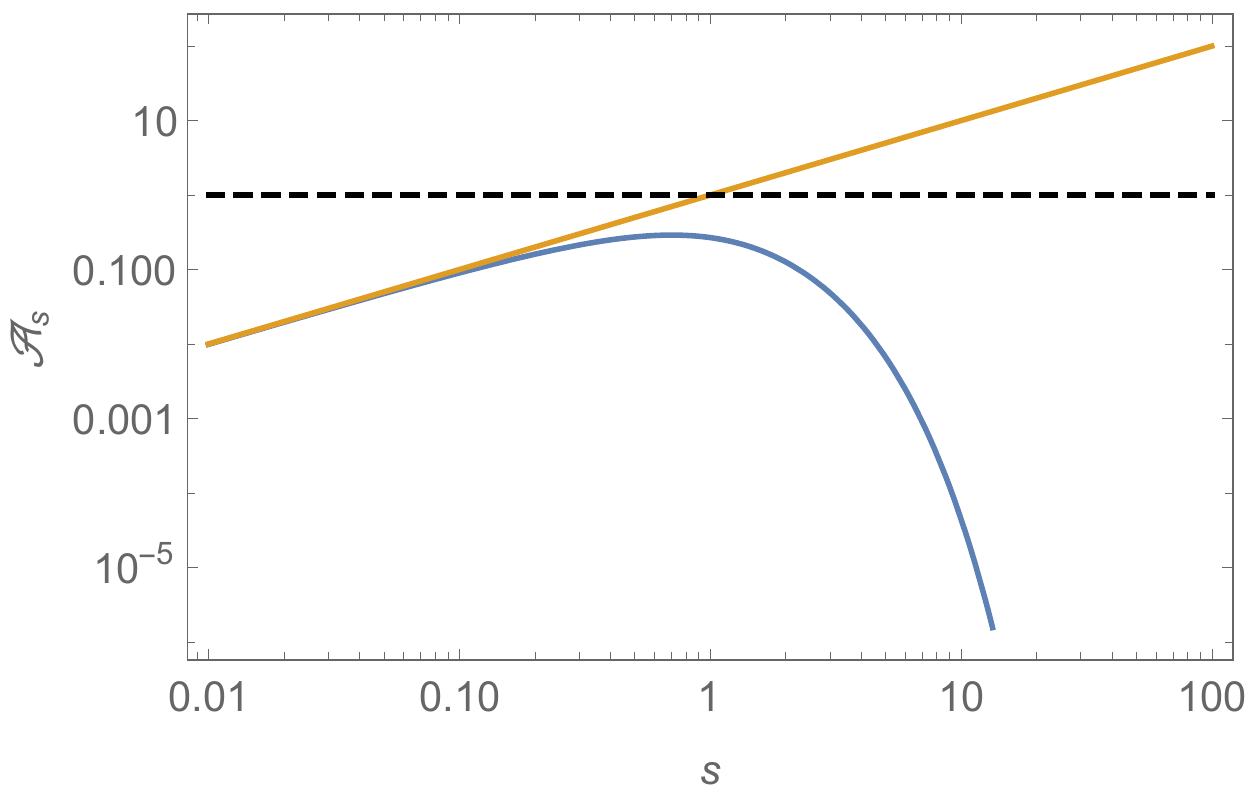}    
	\caption{\label{Fig.amp} The $s$-channel amplitudes obtained from the tanh-model \eqref{eq:fftanh2} (left diagram) and the non-local model \eqref{eq:ffnon-local} (right diagram) related to the $\phi\phi \rightarrow \phi\phi$ scattering process for $\lambda^2 =1$. The straight orange line illustrates the amplitude obtained from a standard kinetic term without form factor. For $s \ll \Lambda^2$ the form factor models agree with this result, while at large centre-of-mass energy, they remove the divergences in the amplitude.}
\end{figure}

Introducing $s \equiv (p_1 + p_2)^2$ as the standard Mandelstam variable associated with the centre-of-mass energy, and using standard Feynman diagram techniques, it is then straightforward to compute the $\phi\phi \rightarrow \phi\phi$ scattering amplitudes resulting from the two choices of form factors
\begin{equation}\label{eq:scalaramp}
\mathcal{A}^{\rm tanh}_s = \frac{\lambda^2 \, s}{1 + s \, \tanh(s)} \, , \qquad
\mathcal{A}^{\rm nl}_s = \lambda^2 \, s \,  {\rm e}^{-s} \, .
\end{equation}
The contributions from the $t$- and $u$-channels have the same structure, and are obtained by crossing symmetry via the replacements $s \leftrightarrow t$ and $s \leftrightarrow u$, respectively.  These amplitudes are illustrated in Fig.\ \ref{Fig.amp}. Their characteristic difference is that $\mathcal{A}^{\rm tanh}_s$ approaches a constant as $s \rightarrow \infty$ while $\mathcal{A}^{\rm nl}_s$ decreases exponentially for large values $s$. Without the form factor contribution, the amplitude would diverge for large centre-of-mass energy (straight line in Fig.\ \ref{Fig.amp}). This divergence is compensated by the form factor, so that the amplitude remains finite as $s \rightarrow \infty$. 

At this stage, it is instructive to consider the momentum-dependent field redefinition
\begin{equation}\label{eq:fieldredef}
\tilde{\phi}(p) \equiv Z(p)^{1/2} \, \phi(p) \, ,
\end{equation}
 with $Z(p)$ being a momentum-dependent wave-function renormalisation. By definition, $Z(p)$ must be \emph{positive and invertible} \cite{tHooft:1973wag}. Since the factor $x$ is not invertible due to its root at $x=0$, the examples \eqref{eq:fftanh2} and \eqref{eq:ffnon-local} suggest to identify $Z(p)$ with the terms given in the straight brackets
\begin{equation}
Z^{\rm tanh}(p^2) =  1 + p^2 \tanh(p^2) \, , \qquad  Z^{\rm nl}(p^2) =  {\rm e}^{p^2}  \, . 
\end{equation}
Rewriting the action \eqref{non-locffaction} in terms of the new field $\tilde{\phi}$ then gives
\begin{equation}
\Gamma[\tilde{\phi}] = \frac{1}{2} \int_x \tilde{\phi} (-\p^2) \tilde{\phi} + \frac{\lambda}{2} \int_x \left(\partial^2 Z^{-\frac{1}{2}}(-\partial^2) \tilde\phi\right) \left(Z^{-\frac{1}{2}}(-\partial^2) \tilde\phi\right)^2 \, .
\end{equation}
Notably, the kinetic term is now given by the standard kinetic term for a real scalar field. Thus the propagator admits a \emph{standard K\"allen-Lehmann representation for a massless field}. Consequently, it is clear that $\tilde{\phi}$ propagates causally albeit being subject to non-local interactions. Since amplitudes are invariant under field redefinitions, one again recovers \eqref{eq:scalaramp} from this action.

At this stage, it is instructive to write the field redefinition eq.\ \eqref{eq:fieldredef} in terms of integral kernels in position space. Converting to Euclidean signature, a form factor satisfying suitable fall-off conditions \cite{Tomboulis:2015gfa} may be represented as a non-local integral kernel
\begin{equation}\label{eq:intkerneldef}
 \phi(x) = Z^{-\frac{1}{2}}(-\partial^2) \tilde\phi(x) = \int \text{d}^4y \, F(x-y) \, \tilde\phi(y) \, .
\end{equation}
The integration kernel $F(x-y)$ has the Fourier representation
\begin{equation}
F(x-y) \equiv \int \frac{\text{d}^4 p}{(2\pi)^4} \, Z^{-\frac{1}{2}}(p^2) \, e^{-i p \cdot (x-y)} \, . 
\end{equation}
The angular integral can be carried out analytically,
\begin{equation}\label{eq:intintegralkernel}
	F(r) = \frac{1}{(2\pi)^2} \frac{1}{r} \int_0^\infty \text{d}p \, p^2 \, Z^{-\frac{1}{2}}(p^2) \, J_1(p \, r) \, . 
\end{equation}
Here $J_\nu(x)$ is the Bessel function of the first kind of order $\nu$ and $r \equiv |x-y|$ is the Euclidean distance between the points $x$ and $y$. For the field redefinition of the non-local gravity model, the integral can then be carried out analytically and is given by a Gaussian,
\begin{equation}\label{eq:kernelnl}
F^{\rm nl}(r) = \frac{1}{4 \pi^2} e^{-r^2/2} \, .
\end{equation}
Thus, in this case, the field redefinition suggests an interpretation in terms of a quasi-local coarse-graining of the scalar field with weight given by the Gaussian \eqref{eq:kernelnl}.

\begin{figure}[t!]
	\centering
	\includegraphics[width = 0.48\textwidth]{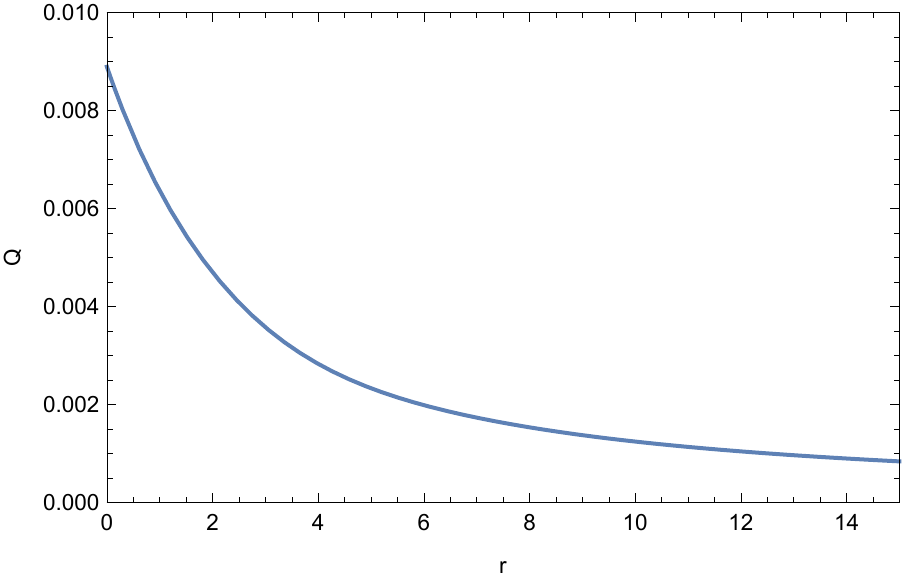} \;
	\caption{\label{Fig.kernels} Illustration of the function $Q(r)$, arising from the $\tanh$-model \eqref{eq:fftanh2}.}
\end{figure}

For the tanh model, one can show that the analogous function is
\begin{equation}\label{eq:Ftanh}
 F^{\tanh}(r) = \frac{1}{4\pi^2r} \left( \frac{1}{r^2} - \frac{1}{2} \right) + Q(r) \, ,
\end{equation}
where $Q(r)$ is a bounded function given by
\begin{equation}
 Q(r) = \frac{1}{4\pi^2r} \int_0^\infty \text{d}p \, p \, J_1(p \, r) \left[ \sqrt{\frac{p^2}{1+p^2 \tanh p^2}} - 1 + \frac{1}{2p^2} \right] \, .
\end{equation}
It is shown in Fig.\ \ref{Fig.kernels}. For large arguments, we find the asymptotic expansion
\begin{equation}
 Q(r) \sim -\frac{1}{4\pi^2r} \left( \frac{1}{r^2} - \frac{1}{2} \right) \, , \qquad \text{as } r \to \infty \, .
\end{equation}
A numerical investigation suggests that this is exact up to terms falling off faster than any power law. Together with the other terms in \eqref{eq:Ftanh}, this would imply that similar to the non-local model, $F^{\tanh}$ is quasi-local in the sense that both integral kernels are exponentially suppressed as $r\rightarrow \infty$. The main difference lies in the divergence of $F^{\tanh}$ for small $r$. This pole is cancelled by the measure term in \eqref{eq:intkerneldef}, thus not giving rise to any  obstacles.

We close our discussion with the following remark. For scalar field theories, it is straightforward to implement field redefinitions of the form \eqref{eq:fieldredef}. For gravity, where the action is invariant under diffeomorphisms, this is less straightforward for two reasons. Firstly, in the gravitational context, the metric $g_{\mu\nu}$ plays the double role of carrying the degrees of freedom of the theory, while at the same time providing the stage on which physics takes place. Secondly, the form factors in \eqref{eq:curvatureexp} generate contributions to both the kinetic as well as to the interaction terms. In order to push the analogy to the scalar model, one could then resort to the background field method, decomposing the spacetime metric $g_{\mu\nu} = \bar{g}_{\mu\nu} + h_{\mu\nu}$ into a fixed background metric $\bar{g}_{\mu\nu}$ and fluctuations $h_{\mu\nu}$ with the latter being ``gravitons'' moving in the background spacetime $\bar{g}_{\mu\nu}$. Performing a momentum-dependent field redefinition for the gravitons, $\tilde{h}_{\mu\nu} = Z(-\bar{D}^2)^{1/2} h_{\mu\nu}$ with $\bar{D}$ the covariant derivative constructed from $\bar{g}_{\mu\nu}$, will again not affect the amplitudes.

\section{Summary and Conclusions}
\label{sect.conclusions}
Studying the options of obtaining a consistent high-energy completion of gravity by including the contribution of form factors in the gravitational dynamics constitutes a promising path towards quantum gravity. Traditionally, non-local gravity programmes make a choice on the form factors \emph{at the level of the bare action}. The task is then to show that this choice results in a perturbatively (super-)renormalisable quantum field theory compatible with other physics requirements as, e.g., a unitary time evolution. Conceptually, this approach differs substantially from the role of the form factors played within \asg{}. In this case, the corresponding functions are not ``put in by hand'', but rather arise as a prediction from the underlying Reuter fixed point. Their low-energy part is determined in terms of the renormalisation group trajectories emanating from the fixed point. Thus the form factors compatible with \asg{} are fixed in terms of the relevant parameters of the fixed point. By now there are several works aiming at determining form factors from first principle computations based on the \FRGE{} \eqref{eq:FRGE}~\cite{Bosma:2019aiu, Bonanno:2021squ, Knorr:2021niv} or from reverse-engineering Monte Carlo data~\cite{Knorr:2018kog}. Owed to the complicated nature of the integro-differential equations encoding these structures, computing the corresponding correlation functions is a highly non-trivial task. Any information on the structure of these functions will provide important clues for solving these equations. In this light, the $\tanh$-model discussed in Sects.\ \ref{sect.2} and \ref{sect.3} serves multiple roles: firstly, it acts as a proof of principle that the Lorentzian quantum effective action leaves sufficient room for rendering \emph{all} amplitudes finite at high energy, thereby realising the core feature of Lorentzian Asymptotic Safety. Secondly, it unravels the ingredients necessary in this construction: an interplay between the high-energy behaviour of propagators and vertices which, in a first-principle derivation, must be provided by the underlying renormalisation group fixed point. Thirdly, the model serves as a catalyst for sharpening questions related to fundamental properties of \asg{} which the construction will have to answer at some point: ``Is the resulting theory causal?'', ``Is there an intuitive physics picture underlying the existence of the non-Gaussian fixed point in terms of non-perturbative degrees of freedom?'', or ``how to formulate a well-defined initial value problem based on the quantum effective action?'' We believe that it is this set of questions where the scientific exchange between non-local, ghost-free gravity and \asg{} can be the most fruitful. The form factors may then constitute the common language for this discussion.

\acknowledgments
We thank L.\ Buoninfante and K.\ Sravan Kumar for organising the workshop \emph{Quantum Gravity, Higher Derivatives \& Nonlocality}, and providing a stimulating and open discussion environment. B.\ K.\ acknowledges support by Perimeter Institute for Theoretical Physics. Research at Perimeter Institute is supported in part by the Government of Canada through the Department of Innovation, Science and Economic Development Canada and by the Province of Ontario
through the Ministry of Colleges and Universities.

\end{document}